\documentclass[aps,prl,twocolumn,showpacs,superscriptaddress]{revtex4}

\usepackage{graphicx}
\usepackage{epsfig}
\usepackage{amssymb}
\usepackage{amsmath}
\usepackage{psfrag}
\usepackage{color}
\usepackage{soul}

\newcommand{\bE}{{\bf E}}
\newcommand{\br}{{\bf r}}

\newcommand{\bG}{{\bf G}}
\newcommand{\dd}{{\mathrm{d}}}
\newcommand{\im}{\mathrm{Im} \, }
\newcommand{\tr}{\mathrm{Tr} \, }

\setstcolor{red}

\begin{document}

\title{Mapping the radiative and the apparent non-radiative local density of states \\ in the near field of a metallic nanoantenna}
\author{D. Cao}
\affiliation{ESPCI ParisTech, PSL Research University, CNRS, Institut Langevin, 1 rue Jussieu, F-75005, Paris, France}
\author{A. Caz\'{e}}
\affiliation{ESPCI ParisTech, PSL Research University, CNRS, Institut Langevin, 1 rue Jussieu, F-75005, Paris, France}
\author{M. Calabrese}
\affiliation{ESPCI ParisTech, PSL Research University, CNRS, Institut Langevin, 1 rue Jussieu, F-75005, Paris, France}
\author{R. Pierrat}
\affiliation{ESPCI ParisTech, PSL Research University, CNRS, Institut Langevin, 1 rue Jussieu, F-75005, Paris, France}
\author{N. Bardou}
\affiliation{Laboratoire de Photonique et Nanostructures (LPN-CNRS), Route de Nozay, 91460 Marcoussis, France}
\author{S. Collin}
\affiliation{Laboratoire de Photonique et Nanostructures (LPN-CNRS), Route de Nozay, 91460 Marcoussis, France}
\author{R. Carminati}
\affiliation{ESPCI ParisTech, PSL Research University, CNRS, Institut Langevin, 1 rue Jussieu, F-75005, Paris, France}
\author{V. Krachmalnicoff}
\email{valentina.krachmalnicoff@espci.fr}
\affiliation{ESPCI ParisTech, PSL Research University, CNRS, Institut Langevin, 1 rue Jussieu, F-75005, Paris, France}
\author{Y. De Wilde}
\affiliation{ESPCI ParisTech, PSL Research University, CNRS, Institut Langevin, 1 rue Jussieu, F-75005, Paris, France}

\begin{abstract}
We present a novel method to extract the various contributions to the photonic local density of states from near-field fluorescence maps. The approach is based on the simultaneous mapping of the fluorescence intensity and decay rate, and on the rigorous application of the reciprocity theorem. It allows us to separate the contributions of the radiative and the apparent non-radiative local density of states to the change in the decay rate. The apparent non-radiative contribution accounts for losses due to radiation out of the detection solid angle and to absorption in the environment. Data analysis relies on a new analytical calculation, and does not require the use of numerical simulations. One of the most relevant applications of the method is the characterization of nanostructures aimed at maximizing the number of photons emitted in the detection solid angle, which is a crucial issue in modern nanophotonics. 
\end{abstract}


\maketitle


Tailoring light-matter interaction is a key issue in modern photonics. A full control of such interaction on the nanometer scale can have a huge impact in a wide range of domains, going from fundamental physics (e.g. control of spontaneous emission with optical antennas~\cite{Vahid2006,Novotny2006,Frimmer2013_superemitter},
strong coupling~\cite{Chang2006, Chang2007,Guebrou2012,Vasa2013}, cavity quantum electrodynamics with localized modes~\cite{Imamoglu, LSapienza2010,Caze2013}) to the design of novel devices (e.g. light harvesting~\cite{Schuller2010, Atwater2010}, photon detection~\cite{Konstantatos2010}, biological sensing~\cite{sensing}). A measurement of light-matter coupling is given by the Purcell factor, which describes the enhancement of the spontaneous decay rate of an emitter in a given environment.  It has been known since the pioneering work by Drexhage~\cite{Drexhage1968} that the decay rate is modified in the vicinity of a metallic structure. However, 
one can observe an enhancement of either the radiative decay rate (corresponding to photon emission in the far field) or the non-radiative decay rate (which measures the coupling to dark modes and absorption losses)\cite{Carminati2006}. Depending on the targeted application (e.g. the design of an efficient single photon source or an efficient quencher of molecular fluorescence), it can be interesting to enhance either the radiative or the non-radiative decay rate. In many situations, the design of a nanostructured medium is aimed at maximizing the number of photons emitted in the far field, in the detection solid angle \cite{Kinkhabwala_2009}. Therefore, controlling and measuring this parameter is a crucial issue in the characterization of nanostructures dedicated to the control of light-matter interaction.
Despite the broad interest, such a measurement has remained challenging and only a few experiments have been reported~\cite{Hecht_PRL, Dahan_PRL, Sandoghdar_PRL}, that only partially address the issue.

In this Letter we present the mapping of the radiative decay rate using a fluorescent nanosource scanned in the near field of a metallic antenna. This is made possible by means of a novel analysis based on the reciprocity theorem. The fluorescent scanning probe allows us to map simultaneously the fluorescence intensity and decay rate, which is proportional to the local density of states (LDOS)~\cite{Krachmalnicoff_OE}, in a confocal geometry in which excitation and detection paths coincide exactly. This allows us to apply rigorously the reciprocity theorem to separate the radiative decay rate from an \emph{apparent} non-radiative decay rate including the contribution of photons radiated out of the detection solid angle and absorption losses, as will be shown by an analytical calculation. The method properly accounts for the radiation pattern of the observed antenna. In order to check the validity of the proposed method, we compare the results obtained from the measurements with a numerical simulation.
   
\begin{figure}[hb!]	
	\begin{center}
			\includegraphics[width=7.5cm]{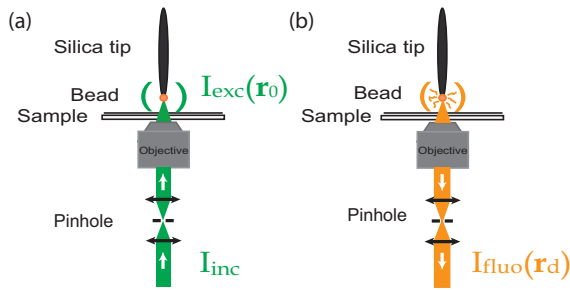}
   \caption{Simplified sketch of the experimental setup. (a) Excitation path. (b) Detection path.}\label{fig_1}
	\end{center}
\end{figure}

The fluorescence intensity emitted by a dipole located in the near field of a metallic nanostructure depends on two processes. $(1)$~The change of the exciting field 
at the position of the emitter induced by the local environment. $(2)$ The balance between the radiative and the non-radiative decay rates, that also depends on the environment. For a dipole located at $\textbf{r}_{0}$, the fluorescence intensity can be written as: 
\begin{equation}\label{eq_fluorescence_signal}
I_{fluo}(\textbf{r}_d)= A\, \frac{\Gamma_{1,\Omega}^{R}}{\Gamma_1}\, \sigma_{abs} I_{exc}(\textbf{r}_{0})
\end{equation}
where $\textbf{r}_d$ is the position of the detector, $\Gamma_1$ is the total decay rate, $\Gamma_{1,\Omega}^{R}=\int_{\Omega_d} \Gamma_1^{R}(\textbf{u}) \, \mathrm{d}\Omega_d$ is the radiative decay rate integrated over the detection solid angle $\Omega_d$ ($\textbf{u}$ being a unit vector defining a detection direction), $I_{exc}(\textbf{r}_{0})$ is the local excitation intensity and $\sigma_{abs}$ is the absorption cross-section. $A$ is a constant that takes into account the detection efficiency and photon losses along the optical path to the detector. 
In the exact confocal geometry considered in this work, exciting and detected photons follow the same path (see Fig.~\ref{fig_1}).
The reciprocity theorem~\cite{Carminati1998, Novotny2011} states that $\boldsymbol{\mu}_1 \cdot \textbf{E}_{exc}(\textbf{r}_{0})=\boldsymbol{\mu}_d \cdot \textbf{E}_{fluo}(\textbf{r}_{d})$
where $\boldsymbol{\mu}_1$ is the dipole moment of the emitter, $\boldsymbol{\mu}_d$ is a (virtual) dipole defining a polarization direction at the detector,
$\textbf{E}_{exc}(\textbf{r}_{0})$ is the exciting field and $\textbf{E}_{fluo}(\textbf{r}_{d})$ is the field reaching the detector.
Since the radiative decay rate is proportional to the power radiated by a classical dipole with dipole moment $\boldsymbol{\mu}_1$, we have for a polarized detection along $\boldsymbol{\mu}_d$:
\begin{equation}\label{eq_reciprocity}
\Gamma^{R}_{1,\Omega}=\tilde{B}|\boldsymbol{\mu}_d \cdot \textbf{E}_{fluo}(\textbf{r}_{d})|^2=\tilde{B}|\boldsymbol{\mu}_1 \cdot \textbf{E}_{exc}(\textbf{r}_{0})|^2=B I_{exc}(\textbf{r}_0)
\end{equation}
where $\tilde{B}$ is a constant and $B=\mu_1 \tilde{B}$. Note that Eq.~(\ref{eq_reciprocity}) is valid at a specific wavelength. In practice, this is a constraint of the applicability of the reciprocity theorem which can be satisfied provided that the response of the nanoantenna is broadband compared to the Stokes shift of the molecule. Inserting Eq.~(\ref{eq_reciprocity}) into Eq.~(\ref{eq_fluorescence_signal}), and summing over all possible dipole orientations $\textbf{u}_1$ and field 
polarizations $\textbf{u}_d$, we end up with an expression for the total detected fluorescence intensity:
\begin{equation}\label{I_fluo_tot}
I^{tot}_{fluo}(\textbf{r}_d)=\frac{A}{B} \sigma_{abs}\left[\frac{\left(\Gamma_{1,\Omega}^{R}\right)^{2}}{\Gamma_{1}}+\frac{\left(\Gamma_{2,\Omega}^{R}\right)^{2}}{\Gamma_{2}}+\frac{\left(\Gamma_{3,\Omega}^{R}\right)^{2}}{\Gamma_{3}}\right]
\end{equation}
where the subscripts $(1,2,3)$ refer to three orthogonal directions of the transition dipole.
This expression directly connects the detected fluorescence intensity to the total and the radiative decay rate integrated over the detection solid angle. It is an essential relationship in the method described in this Letter, as we shall see in the analysis of the experimental data.


Equation \ref{I_fluo_tot} can be applied to data analysis provided that the setup satisfies the following crucial points: i) the excitation and the detection paths have to be reciprocal; ii) the fluorescence intensity and the decay rate have to be measured with the same fluorescent emitter. These two requirements are verified in our recently developed experimental setup, which is a fluorescent near-field scanning probe microscope \cite{Krachmalnicoff_OE, Frimmer2011} modified in such a way that the excitation and detection paths are rigorously the same, as sketched in Fig.\ref{fig_1}.

The sample is mounted on a sample-scanning inverted confocal microscope combined with a homebuilt atomic force microscope (AFM). The AFM tip is a tapered optical fiber. A fluorescent bead (Invitrogen Red Fluospheres, diameter 100~nm) is grafted at the extremity of the tip of the AFM. The excitation is performed at 560~nm through a 150~$\mu$m confocality pinhole and an oil objective (N.A. 1.4) with a supercontinuum pulsed laser (Fianium SC450) at a repetition rate of 10~MHz. Fluorescence photons are collected through the same objective, pass in the confocality pinhole and then are separated from the exciting photons with a dichroic mirror and a high-pass filter ($\lambda > $~594~nm). 
Importantly, the use of the same confocal pinhole on the excitation and detection optical paths, ensures that excitation/detection photons are emitted/detected through the same optical mode. This is crucial for the application of the reciprocity theorem. 

Time-resolved photon detection is performed with a time correlated single photon counting system (MPD PDM-series avalanche photodiodes combined with Picoquant HydraHarp 400 acquisition board), which allows one to simultaneously map the fluorescence intensity and decay rate.
The fluorescent scanning probe is held by shear force feedback at a constant distance of approximately $20$~nm to the surface of a nanostructured sample while the latter is scanned. The topography of the sample is recorded simultaneously with the total decay rate $\Gamma$ and fluorescence intensity maps.

We present the study of the response of a single gold nanodisc, 130~nm in diameter and 30~nm thick, obtained by electron beam lithography on a glass substrate.
\begin{figure}[ht!]	
	\begin{center}
			\includegraphics[width=7.5cm]{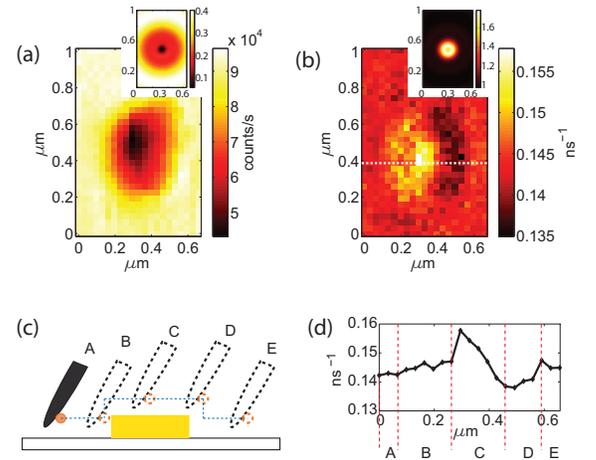}
   \caption{\label{fig_2} (a) Fluorescence intensity map for one gold nanodisc. (b) Decay rate map for the same gold nanodisc. Insets: numerically calculated maps. The decay rate map has been normalised by its value in vacuum. (c) Path followed by the fluorescent probe during the scan (not to scale). (d) Cross section through the row indicated with a white dashed line in~(b).}
	\end{center}
\end{figure}

Figures \ref{fig_2}(a,b) show the fluorescence intensity and decay rate maps for the gold monomer. The intensity map has been corrected for a slow progressive lateral drift of the fluorescent probe with respect to the exciting laser beam (on the order of $3$~nm per minute), which results in a progressive decrease of the mean intensity detected during the scan. We therefore start the data analysis process by subtracting a gradient background from the intensity map and by adding an offset corresponding to the mean value of the background. This processing does not affect the LDOS map.          
As confirmed by numerical simulations shown in the insets in Fig.~\ref{fig_2}, we observe an increase of the LDOS and a decrease of the fluorescence intensity in correspondence with the metallic structure. The contrast of the numerical maps is overestimated due to the absence of the substrate in the simulation. The experimental LDOS map shows a zone where the LDOS has a smaller value than on the glass substrate, and a zone where it has an intermediate value between that on the glass substrate and that on the gold disc. This behaviour is due to a grafting of the fluorescent bead on the side of the silica tip, as depicted in Fig.~\ref{fig_2}(c). Figure~\ref{fig_2}(d) shows the profile of the decay rate along the white dashed line in Fig.~\ref{fig_2}(b). At the beginning of the scan (zone A in Fig.~\ref{fig_2}(c,d)), both the tip and the bead are at a given height on the glass coverslip. As the tip approaches the nanodisc (zone B), the bead gets closer to the gold disc and the LDOS slowly increases. Then both the tip and the bead are scanned on top of the disc and the bead feels an enhanced LDOS (zone C). Approaching the disc edge, the tip remains on the disc, while the bead is driven out of it (zone D). In this position the bead feels a reduced LDOS compared to that on the glass coverslip because of the larger glass-bead distance (on the order of $50$~nm). Finally, both the tip and the bead are scanned over the glass coverslip (zone E) and the bead feels the same LDOS as in the initial position. Note that we have checked that the value of the decay rate measured in zone D is the same as that measured when the tip is retracted by 30~nm above the bare glass coverslip.

In the following, we describe the procedure used to extract, from the experimental data, the maps of the different contributions to the decay rate, which include radiated photons towards the detector and losses by radiation out of the detection solid angle or by absorption. Since the fluorescent bead contains several thousands molecules with random orientations, we are not able to measure the ratio $\Gamma_{i,\Omega}^{R}/ \Gamma_{i}$ in the three directions of the transition dipole. We  introduce an LDOS anisotropy factor $C_i=\Gamma / \Gamma_i$ and assume an isotropic response of the system ($C_{1,2,3}=C$). The relevance of such an assumption will be checked
afterward by comparison with numerical simulations. In these conditions, Eq.~(\ref{I_fluo_tot}) can be rewritten as:
\begin{equation}\label{eq_fluo_fin}
I^{tot}_{fluo}(\textbf{r}_d)=3\, \frac{A C}{B} \,\sigma_{abs}\frac{(\Gamma_{rms,\Omega}^{R})^2}{\Gamma}
\end{equation}
where 
\begin{equation}
\Gamma_{rms,\Omega}^{R}=\left[ \frac{\left(\Gamma_{1,\Omega}^{R}\right)^{2}
+\left(\Gamma_{2,\Omega}^{R}\right)^{2}+\left(\Gamma_{3,\Omega}^{R}\right)^{2}}{3} \right]^{1/2}
\end{equation}
is the effective radiative decay rate that we extract from the experimental data. The parameter $\Gamma_{rms,\Omega}^{R}$ is representative of the fraction of photons radiated in the far field in the detection solid angle, averaged over the three dipole orientations $i$.
In order to deduce $\Gamma_{rms,\Omega}^{R}$ from the measurement of $I^{tot}_{fluo}(\textbf{r}_d)$ and the total decay rate $\Gamma$, we have to get rid of the unknown prefactor in Eq.~(\ref{eq_fluo_fin}).  
This is done by measuring the fluorescence intensity and the decay rate with the same bead, in the same experimental conditions as for the measurement of the antenna, but on a bare glass substrate used as a reference sample. In practice, we use the values of the intensity and decay rate on the first pixel of the maps (i.e. far from the gold structure). Since the non-radiative component of the decay rate is negligible on this reference sample, the reference fluorescence intensity is given by 
$I^{ref}_{fluo}(\textbf{r}_d)=3\, (AC/B) \,(\Omega/4\pi)^2 \sigma_{abs}\, \Gamma_{ref}$. This allows us to rewrite Eq.~(\ref{eq_fluo_fin}) in the form
\begin{equation}\label{eq_fluo_fin_ref}
I^{tot}_{fluo}(\textbf{r}_d)=\left(\frac{4\pi}{\Omega}\right)^2 \,\frac{I^{ref}_{fluo}(\textbf{r}_d)}{\Gamma_{ref}}\,\frac{(\Gamma_{rms,\Omega}^{R})^2}{\Gamma} .
\end{equation}
Note that we consider that the constant $C$ is the same on the reference and the real sample, an assumption whose relevance will also be assessed through the comparison with numerical simulations.

\begin{figure}[ht!]	
	\begin{center}
			\includegraphics[width=7.5cm]{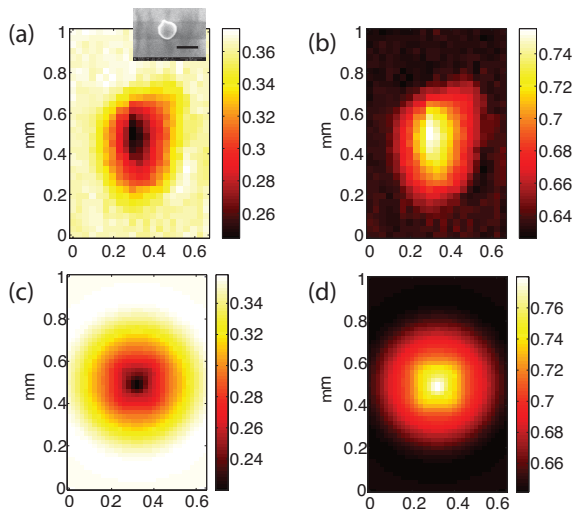}
   \caption{\label{fig_3}(a,b) Measured effective radiative decay rate $\Gamma_{rms,\Omega}^{R}$ and \emph{apparent} non-radiative decay rate $\tilde \Gamma_{rms,\Omega}^{NR}$ map for a single gold nanodisc, normalized by the measured total decay rate map. Inset: SEM image of the sample, black bar: $200$~nm. (c,d) Numerical maps of the effective radiative decay rate $\Gamma_{rms,\Omega}^{R}$ and the \emph{apparent} non-radiative decay rate $\tilde\Gamma_{rms,\Omega}^{NR}$ normalized by the calculated total decay rate map.}
	\end{center}
\end{figure}
	
The map of the effective radiative decay rate $\Gamma_{rms,\Omega}^{R}$ in the near field of the gold monomer, as deduced from the experimental data, is shown in Fig.~\ref{fig_3}(a).  A measurement of an \emph{apparent} non-radiative decay rate, including photons which are not detected either because they are radiated out of the detection solid angle ($\Gamma_{rms,4\pi-\Omega}^{R}$) or because they are absorbed by the nanostructure ($\Gamma_{rms,\Omega}^{NR}$), is given by $\tilde\Gamma_{rms,\Omega}^{NR}=\Gamma-\Gamma_{rms,\Omega}^{R}=\Gamma_{rms,\Omega}^{NR}+\Gamma_{rms,4\pi-\Omega}^{R}$. The map of $\tilde \Gamma_{rms,\Omega}^{NR}$ is shown in Fig.~\ref{fig_3}b.
We note that the contribution of photons radiated out of the detection solid angle could be removed by integrating the measurement of the fluorescence intensity and the decay rate over all the full solid angle of emission. Therefore the method presented in this Letter can be used to extract both the radiative and the non-radiative decay rates, provided that the photon detection in the far field is performed over the full solid angle. In Fig.\ref{fig_3}, both decay rate maps have been normalized by the total decay rate map shown in Fig. \ref{fig_2}(b). This normalization allows us to 
put forward the change in the LDOS due to the nanostructure itself, and to get rid of the contribution of the substrate, of the exact probe shape and tip trajectory, as confirmed by comparison to numerical simulations presented below. The non-normalized data are shown in the Supplemental Material. Note that the normalized effective radiative decay rate is the relevant quantity to be maximized in an experiment aimed at detecting the emission of as many photons as possible from a fluorophore coupled to an optical antenna or to any other nanostructured material, which is crucial in many practical cases.

In order to get insight into the experimental results, and to assess the validity of the two hypotheses made in the procedure for the data analysis, we have performed numerical simulations using the method described in Refs.~\cite{Krachmalnicoff_OE,Caze12}. We stress that the determination of $\Gamma_{rms,\Omega}^{R}$ and $\tilde \Gamma_{rms,\Omega}^{NR}$ from the experimental data does not require the use of numerical simulations, which is a strength of the procedure presented above. The calculation of the electric field is based on the volume integral equation
\begin{equation}
\label{volume_integral_equation}
\bE(\br) = \bE_0(\br) + k_0^2\int_V \left[\epsilon(\omega) - 1\right] \bG_0(\br,\br',\omega) \bE(\br^\prime) \dd \br^\prime,
\end{equation}
where $k_0=\omega/c$, with $c$ the speed of light in vacuum, $V$ is the volume occupied by gold, $\bE_0$ is the incident field, $\bG_0$ is the dyadic Green function of the host medium (free space in the simulations performed here) and $\epsilon(\omega)$ is the dielectric function of gold, taken from Ref.~\cite{PalikBook}.
Equation~(\ref{volume_integral_equation}) is solved by a moment method without any approximation~\cite{HarringtonBook}, where the 
Green function $\bG_0$ is integrated over the discretization cells ($2.5$~nm cubic cells) to improve convergence~\cite{Chaumet04}. 
Solving Eq.~(\ref{volume_integral_equation}) under illumination by a plane wave (excitation stage) or an electric dipole source (emission stage) allows us to compute 
the total electric field $\bE$ (or equivalently the total Green function $\bG$) everywhere, and to deduce all the parameters entering the fluorescence decay rates and intensity. The total decay rate is deduced from 
$\Gamma(\br_0,\omega)/\Gamma_0 = (2\pi/k_0) \, \im\,\tr\, \left[ \bG(\br_0,\br_0,\omega) \right]$,
where $\Gamma_0$ is the decay rate in vacuum and $\tr$ denotes the trace of a tensor. 
The radiative decay rate $\Gamma_{i,\Omega}^{R}(\omega_{fluo})$ integrated over the detection solid angle, for a given orientation $i$ of the transition dipole,
can be computed from the far-field radiation pattern at the emission frequency $\omega_{fluo}$~\cite{Caze12}.
The excitation intensity $I_{exc}(\textbf{r}_{0})$ is calculated using a plane-wave illumination at normal incidence and at the excitation frequency $\omega_{exc}$. The calculation is averaged over two orthogonal polarizations to mimic the unpolarized laser used in the experiment. Excitation and emission wavelengths are set to  560~nm and 605~nm respectively.

In order to directly compare simulations and experiments, we have calculated the effective radiative decay rate $\Gamma^R_{rms,\Omega}$ and the \emph{apparent} non-radiative decay rate maps $\tilde \Gamma_{rms,\Omega}^{NR}$. The simulated maps, normalized by the calculated total decay rate map, are shown in Fig.~\ref{fig_3}(c,d).
Numerical simulations and experiments are in good quantitative agreement. This proves that the assumptions made in the theoretical model used to extract the effective radiative decay rate from the measurement of the total decay rate and the fluorescence intensity are relevant. Moreover, since the substrate and of the tip are not accounted for in the present simulation,
the quantitative agreement supports the claim that the normalization of the LDOS maps strongly reduces their influence, favoring changes in the LDOS driven by the nanostructure itself.

The resolution and the quality of the experimental maps can be improved by grafting a smaller fluorescent emitter on the extremity of the tip and by controlling the grafting position, or by  using a single quantum emitter with a given dipole orientation \cite{vanHulst_2014}. We also note that the measured maps are likely to change if the excitation and emission wavelengths change and if the detection and excitation directions change. In both cases, photons will populate different modes of the electromagnetic field affecting consequently the measured decay rate maps \cite{Knight10}.

The method described above can be applied to more complex systems. As an example, the measurements of the radiative and the \emph{apparent} non radiative decay rate maps on a nanoantenna formed by a linear chain of three 130~nm diameter gold nanodiscs separated by 20~nm gaps is shown in the Supplemental Material.

In conclusion, we have introduced an experimental method to map the radiative and the \emph{apparent} non-radiative local density of photonic states which is crucial for the study and the characterization of nanostructured  samples for a wide range of applications aimed at maximizing the number of photons radiated in the far field in the detection solid angle. The key point is the simultaneous measurement of the fluorescence intensity and decay rate in an exact confocal geometry, permitting a rigorous use of the reciprocity theorem. In the case of a single gold nanodisc, the experimental procedure is in quantitative agreement with exact numerical simulations, thus proving the relevance of the approach. The general applicability of the method has been demonstrated on an optical antenna but it could also be applied to other metallic or dielectric nanostructures.

We acknowledge Abdel Souilah for technical support. This work was supported by LABEX WIFI (Laboratory of Excellence within the French Program ``Investments for the Future'') under reference ANR-10-IDEX-0001-02 PSL*, by the Region Ile-de-France in the framework of DIM Nano-K and by the French National Research Agency (ANR-11-BS10-0015, ``3DBROM").  


\end{document}